# Deep Learning Captures More Accurate Diffusion Fiber Orientations Distributions than Constrained Spherical Deconvolution

Vishwesh Nath, Kurt G. Schilling, Colin B. Hansen, Prasanna Parvathaneni, Allison E. Hainline, Camilo Bermudez, Andrew J. Plassard, Vaibhav Janve, Yurui Gao, Justin A. Blaber, Iwona Stępniewska, Adam W. Anderson, Bennett A. Landman

**Synopsis**

Confocal histology provides an opportunity to establish intra-voxel fiber orientation distributions that can be used to quantitatively assess the biological relevance of diffusion weighted MRI models, e.g., constrained spherical deconvolution (CSD). Here, we apply deep learning to investigate the potential of single shell diffusion weighted MRI to explain histologically observed fiber orientation distributions (FOD) and compare the derived deep learning model with a leading CSD approach. This study (1) demonstrates that there exists additional information in the diffusion signal that is not currently exploited by CSD, and (2) provides an illustrative data-driven model that makes use of this information.

**Introduction**

Understanding the relationship between observed diffusion weighted MRI signals and true tissue microarchitecture is of fundamental concern for biophysical modeling, detecting microstructural differences, and brain tractography. Substantial efforts have been invested in interpreting the diffusion signal from both model-based (e.g., constrained spherical deconvolution - CSD [1,2], Q-ball [3], persistent angular structure - PAS [4]) and data-driven [5] perspectives. Recently, multi-layer neural networks (or informally, deep learning or deep neural networks - DNN) have emerged as a leading class of machine learning approaches. Moreover, advances combining MRI and whole brain histology have enabled volumetric registration between MRI and histological processes, while co-registered confocal microscopy allows direct 3D observation of intra-voxel tissue orientation. Here, we apply deep learning to investigate the potential information content in single shell diffusion weighted MRI to explain histologically observed fiber orientation distribution (FOD) functions.

**Data**

Three ex-vivo squirrel monkey brains were imaged on a Varian 9.4T scanner. Briefly, data were acquired with a 3D diffusion-weighted EPI sequence (b-value=6,000 s/mm2, 100 directions) at 300um isotropic resolution. After scanning, the tissue was sectioned, stained with the fluorescent DiI, and imaged on an LSM710 Confocal microscope following the procedures outlined in [6]. The histological FOD was extracted using structure tensor analysis. Finally, a multi-step registration procedure [6] was used to determine the corresponding diffusion MRI signal. A total of 567 histological voxels were processed, and a hundred random rotations were applied to each one of them for both the MR signal and the histology FOD to augment the data bringing the total to 57267 voxels [7].

For qualitative validation, a single healthy human volunteer was scanned for a single session using a 3T (Achieva, Philips Medical Systems, Best, The Netherlands) with a 32-channel head coil. Four scans acquired were at a b-value of 2000 s/mm$^2$ (which approximates the diffusion contrast of a fixed ex vivo scan at a b-value of 6000 s/mm$^2$) with 96 gradient directions and an additional b0 per scan (2.5mm isotropic resolution, matrix of 96x96, 38 slices, Multi-Band=2; SENSE=2.2;TR= 2650 ms; TE=94 ms; partial Fourier=0.7). Standard pre-processing with FSL (topup, eddy correction, registration, averaging across scans) was performed before analysis.

**Method**

Both ex-vivo and in-vivo HARDI acquisitions were fit with $8^{th}$ order real spherical harmonics. Outliers were manually reviewed for imaging artifacts, and 54 voxels were removed. FOD's from the histology were fitted with a $10^{th}$ order real spherical harmonics. Histology data was divided into training/validation (44,541 voxels) and testing sets (7,272 voxels) without mixing augmented data (rotations). For training/validation, a 20% percent split was used with 5 folds. Mean squared error was used to assess model accuracy [8].

**Results**

The median angular correlation coefficient (ACC) for CSD (0.7965) was significantly ($p<0.05$, non-parametric signed rank test) lower than for the deep approach (0.8165) (Fig 2), which corresponded to a lower root mean squared error for the deep approach (0.539 versus 0.561). Qualitatively, the predicted FOD's on the human in vivo data demonstrate anatomical consistency (Fig 3), indicating that the deep learning approach is predicting structure in line with prior observations.

**Discussion**

By demonstrating superiority of a deep learning approach over a leading model-based approach, CSD, we show that (1) there exists additional information in the diffusion signal that is not currently exploited by CSD, and (2) provide an illustrative data-driven model that makes use of this information. In a preliminary analysis, we applied the same network to ex vivo imaging at a b-value of 9000 s/mm$^2$ and found a significantly higher ACC (0.850, $p<0.05$, non-parametric signed rank test) for deep learning which is 6.7% higher than CSD. Hence, generalizing the deep learning to use multiple shells and adapt to high b-values is a promising area of exploration. To enable others to investigate our results, the derived TensorFlow models that describe the identified MRI:histology relationships are available on the NITRC project "masimatlab" . Perhaps most importantly, this deep learning analysis demonstrates that current models for identifying fiber orientation distributions do not make all possible use of existing information, and additional innovation is possible. The deep learning models presented herein are preliminary and have not guaranteed optimality properties, and further exploration of the space of multi-layer neural networks is warranted. Additionally, continued refinement of deep learning approaches could make use of both traditional data augmentation of ground truth (e.g., rotations as used herein), but also physics/diffusion simulations of modeled geometry along with image acquisition models.

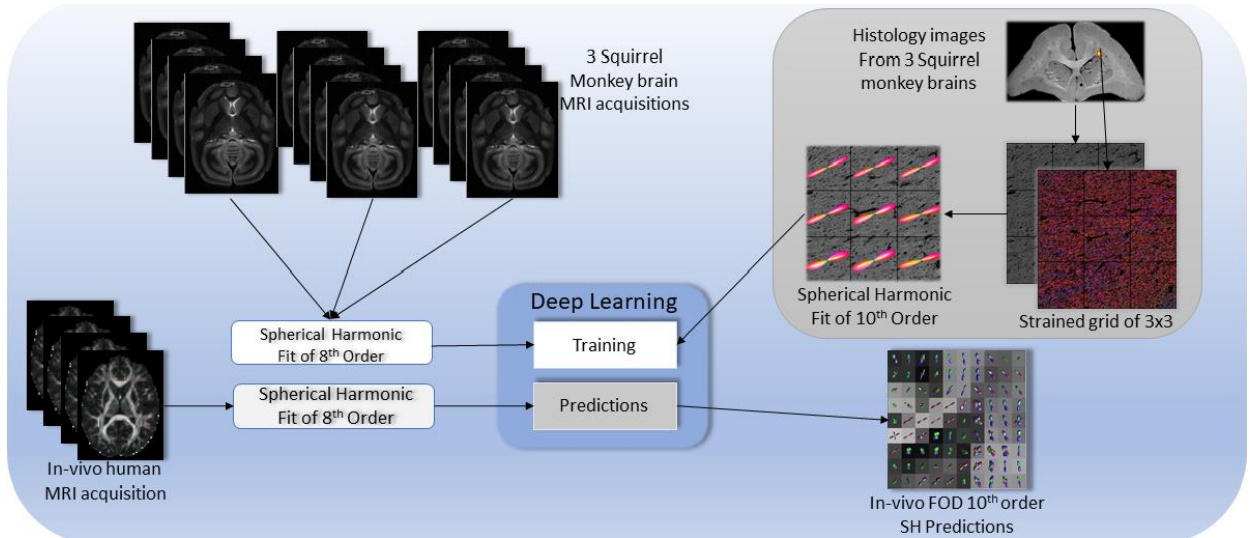

Fig 1. Confocal histological data provides a ground truth basis for fiber orientation distributions. The truth data was split into a training set and a testing set. Once trained, the deep learning approach was applied to both the testing set and a separate human dataset.

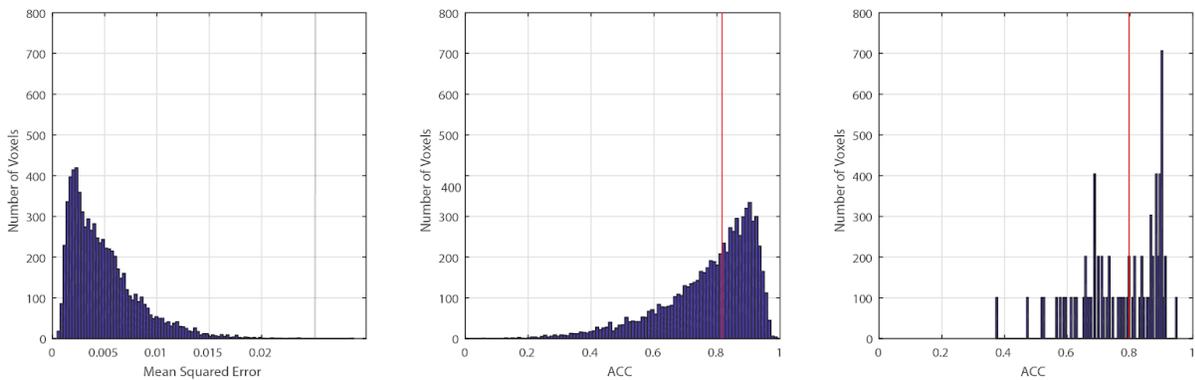

Fig 2. A) Histogram of MSE across all voxels between histology and DNN predicted FOD's. B) Histogram of ACC across all voxels from the test set of histology and DNN predicted FOD's. Media ACC is 0.817 C) Histogram of ACC across all voxels from the test set of histology and CSD predicted FOD's. Median ACC is 0.797

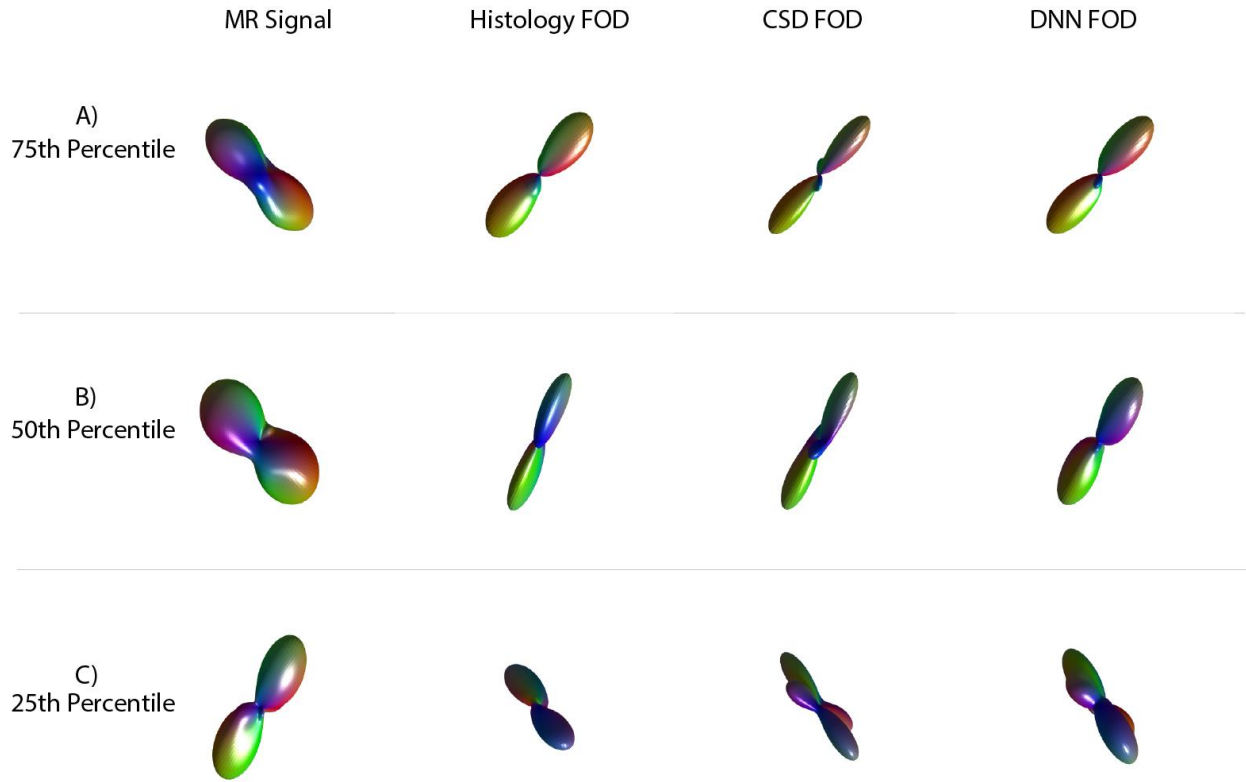

Fig 3. Qualitative visualizations of the MRI fitted to 8th order SH, Histology FOD 10th order SH, CSD 8th order SH, DNN prediction 10th order SH (in order per row). A) 75th percentile (0.936) of ACC for DNN. B) 50th percentile (0.817) of ACC for DNN. C) 25th percentile (0.740) of ACC for DNN.

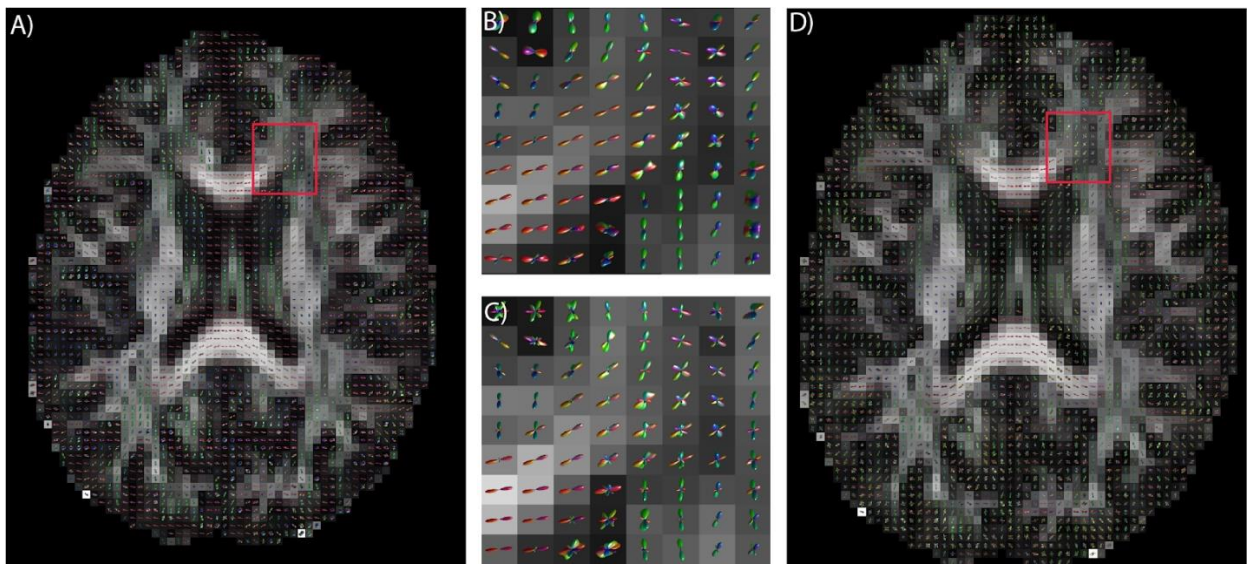

Fig 4. A.) Prediction of deep learning model on human in vivo data at a b-value of 2000 s/mm shown on a middle axial slice. B.) deep learning models predictions zoomed region of interest in the pons of corpus callosum. C) CSD predictions zoomed region of interest in the pons of corpus

callosum. D) Predictions of CSD on human in vivo data at a b-value of 2000 s/mm shown on a middle axial slice.